\documentclass{article}
\font\cero=cmss10 scaled 1728 
\usepackage[inline]{showlabels}
\usepackage{mathtools}
\DeclarePairedDelimiter\ket{\lvert}{\rangle}
\DeclarePairedDelimiterX\braket[2]{\langle}{\rangle}{#1 \delimsize\vert #2}
\usepackage{setspace}
\usepackage{float}
\usepackage{subcaption}
\usepackage{physics}
\usepackage{tabularx} 
\usepackage{tensor}
\usepackage{verbatim}
\usepackage{amsfonts}
\usepackage{graphicx}
\usepackage{amssymb}
\usepackage{float}
\usepackage{comment}
\excludecomment{toexclude}
\begin{document}
\begin{flushleft}
{\cero 
Weighted Lorentz invariant measures as quantum field theory regulators
}\\
\end{flushleft} 
{\sf R. Cartas-Fuentevilla, and A. Mendez-Ugalde }\\
{\it Instituto de F\'{\i}sica, Universidad Aut\'onoma de Puebla,
Apartado postal J-48 72570 Puebla Pue., M\'exico.} \\

ABSTRACT: 

In this work we develop a re-formulation of quantum field theory through the more general weighted Lorentz invariant 
measures that the definition of quantum fields allows; this approach 
provides finite answers for the long-live problems of the traditional formulations of quantum field theories, namely, smooth distributions for the field commutators that are finite a short distances, finite vacuum expectation values for the energy (without invoking normal ordering of operators), and finite fluctuations for the field operators.
Our construction is based on a critical point of view on conventional quantum field theory statements, instead of invoking string theory inspired frameworks, since they are not necessary. We shall show that the conventional scheme for constructing quantum field theories has the necessary ingredients for obtaining  generalized versions that, respecting the Lorentz symmetry, allows us to cure some of the divergences that plague the different formulations, particularly the ultra-violet divergences; additionally the present scheme will allows us to construct an infinite family of noncommutative field theories that are compared with other formulations. At the end, we discuss the impact of our formulation on particle physics numerology and on the cosmological constant problem.\\

\noindent KEYWORDS: noncommutative quantum field theories; ultraviolet divergences; finite vacuum expectation values; finite quantum field theory.

\section{Antecedents, motivations, and results}

The origin of the divergences in QFT can be traced to the construction of the canonical commutation relations for the quantum fields and the requirement of Lorentz symmetry; hence, the creation of a finite quantum field theory would require  the modification of  the former and/or the abandonment of the later. Many different theoretical frameworks have been developed by exploring the Lorentz symmetry breaking as a quantum field theory regulator; additionally this idea is largely motivated by the theoretical predictions coming from quantum gravity and string theory frameworks, namely, that the Lorentz symmetry is not an exact symmetry at high energies (see for example \cite{matt}, and  \cite{mattin}). Furthermore, the canonical commutation relations for quantum fields can be modified 
by the appearance of non-commutative spaces in string theory frameworks, and recent developments in non-commutative quantum mechanics \cite{falomir,aca,nair};
the resulting theories violate relativistic invariance, but they suffer still of ultraviolet divergences (see for example \cite{gamboa}); hence
the problem is far from to be solved.

In this manuscript we discuss whether actually the quantum gravity and/or string theory inspired frameworks are necessary for providing the clues for a finite quantum field theory; rather, we shall show that the conventional scheme may have the seeds for the possible creation of finite quantum field theories; the key observation is that the canonical hypothesis of promoting the commutation relations (of the dynamical variables) in quantum mechanics, to the commutation relations (for the dynamical variables) in quantum field theory, through the transition  to the continuum through $\delta_{ij}\rightarrow \delta_{Dirac}(\vec{x}-\vec{y})$, allows to incorporate
 smooth distributions that will lead to smooth away the singularities that originate the divergences.
As we shall see, the Dirac delta function is only one element (in fact the only divergent one in all dimensions) of an infinite family of distributions available for describing finite quantum field theories,
and thus we shall be able to construct the field commutators in terms of smooth distributions that will maintain the symmetries of the Dirac delta function, but will be also dependent on the background dimension, on the mass, and will be finite at short distances (sections \ref{oneplusone}, \ref{twoone}, and \ref{threeone}).

Once we have regularized the field commutators, the consequence direct is the regularization of the vacuum energy (section \ref{vev}), one of the most famous long-live problems of quantum field theory; our regularization scheme does not require to invoke the normal ordering of operators for extracting the infinite vacuum energy, and thus it is consistent with general relativity. Additionally the vacuum expectation values for field operators, their fluctuations at fixed points, will be finite, as opposed to the traditional belief (section \ref{fluctuation}). 
We finalize the manuscript with some concluding remarks, on some extensions and future developments of  our results.

As preliminary element, we recall that quantum field theories are constructed by imposing the canonical relations used in ordinary quantum mechanics, namely, 
\begin{eqnarray}
[x_{i},p_{j}]=i\delta_{ij},\label{qm1}\\
\big[x_{i},x_{j}]=0, \quad [p_{i},p_{j}]=0, 
\label{qm2}
\end{eqnarray}
on the field commutation relations, by considering the transition to the continuum $\delta_{ij}\rightarrow \delta(\vec{x}-\vec{y})$. For concreteness, for a complex scalar field we have
\begin{eqnarray}
[\hat{\psi}(\vec{x}), \hat{\Pi}_{\psi}(\vec{y})]=i\delta(\vec{x}-\vec{y}); \quad
\label{field}\\
\big[\hat{\psi}(\vec{x}), \hat{\psi}^{\dagger}(\vec{y})\big]=0, \quad
\big[\hat{\Pi}_{\psi}(\vec{x}), \hat{\Pi}_{\psi}^{\dagger}(\vec{y})\big]=0;
\label{field0}\\
\big[\hat{\psi}(\vec{x}), \hat{\Pi}_{\psi}^{\dagger}(\vec{y})\big]=0; \label{field1} 
\end{eqnarray}
where $\hat{\Pi}_{\psi}$ stands for the conjugate momentum for $\psi$; these relations are evaluated {\it at the same time}, at different spatial locations $\vec{x}$, and $\vec{y}$. 
These commutation relations described by Dirac delta functions are 
valid for arbitrary background dimensions, apply indistinctly for the massless or massive case, and are divergent at short distances  $(\vec{x}-\vec{y})\rightarrow 0$; by depending on the difference $(\vec{x}-\vec{y})$,
they are translational invariant and symmetric under $(\vec{x}-\vec{ y})\rightarrow -(\vec{x}-\vec{y})$.
In the next section, we smear the field commutator (\ref{field}) over space, by relaxing the transition to the continuum through $\delta_{ij}\rightarrow D (\vec{x}-\vec{ y})$, where $D(\vec{x}-\vec{ y}) $ is  a smooth distribution that will be obtained by choosing all convergent Lorentz invariant measures in the momenta space that the definition of field operators admits; this simple criterion will allow construct the infinite family of distributions commented above.

\section{Weighted invariant measures}
\label{weight}
We study for simplicity a complex scalar field in $D+1$ dimensions with Lagrangian
${\cal L}= \partial_{\mu}\varphi\partial^{\mu}\overline{\varphi}-m^2\varphi\overline{\varphi}$, with equations of motion given by $(\Box+m^2)\varphi=0$;
 our results with global $U(1)$ symmetry can be generalized to local $U(1)$ gauge symmetry in a direct way. With the decompositions
for the quantum field and its conjugate momentum, 
\begin{eqnarray}
\hat{\varphi}(\vec{x},t)=\frac{1}{\sqrt{(2\pi)^D}}\int \frac{d\vec{k}}{w_{k}^{\frac{s}{2}}}\big[\hat{a}_{k} e^{-iw_{k}t}e^{i\vec{k}\cdot\vec{x}}+\hat{b}_{k}^{\dagger}   e^{iw_{k}t}e^{-i\vec{k}\cdot\vec{x}} \big];
\label{fieldo}\\
\hat{\pi}(\vec{x},t)=\frac{i}{\sqrt{(2\pi)^D}}\int \frac{d\vec{k}}{w_{k}^{\frac{s}{2}-1}}\big[\hat{a}_{k}^{\dagger} e^{iw_{k}t}e^{-i\vec{k}\cdot\vec{x}}-\hat{b}_{k} e^{-iw_{k}t}e^{i\vec{k}\cdot\vec{x}} \big];
\label{fieldo1}
\end{eqnarray}
the equations of motion are satisfied provided that  the usual dispersion relation $w_{k}^2=m^2+\vec{k}^2$ holds; the wight $s$ of the Lotentz invariant measure $\int \frac{d\vec{p}}{w_{k}^{\frac{s}{2}}} $  is an integer (by simplicity); typically with the choice $s=1$, one is enforcing the transition to the continuum with $[\hat{\varphi},\hat{\pi}]\rightarrow \delta(\vec{x}-\vec{y})$; however, we are not bound to satisfy this very restrictive requirement, rather we are obligated to respect the Lorentz invariance, and then we keep the expressions for arbitrary $s$, and we have the convergence as the only criterion for constructing the quantum theories. Furthermore, the (nonvanishing) commutation relations for the annihilation/creation operators read
\begin{eqnarray}
[\hat{a}_{k},\hat{a}_{k'}^{\dagger} ]=\alpha\delta(k-k'),\quad [\hat{b}_{k},\hat{b}_{k'}^{\dagger} ]=\beta\delta(k-k'),
\label{ac}
\end{eqnarray}
where $\alpha$ and $\beta$ are real parameters, and will play a nontrivial role in the approach at hand; the traditional choice $\alpha=1=\beta$ is only a possible election. In fact our present criticism on the textbook statements includes 
the commutation relations between annihilation/creation operators, but we restricted ourselves to the deformation described 
in (\ref{ac}), and we shall develop a more general deformation scheme elsewhere.

With the commutators (\ref{ac}), the general transition will be achieved through $[\hat{\varphi},\hat{\pi}]\rightarrow D_{s}(\vec{x}-\vec{y})$, where $D_{s}$
are smeared versions of the Dirac delta, that will depend on the mass, on the spatial interval, and are finite to short distances; for arbitrary dimension, the nontrivial commutators read
\begin{eqnarray}
    \left[\hat{\varphi}(\vec{x},t),\hat{\pi}(\vec{x'},t) \right]=\frac{i(\alpha+\beta)}{(2\pi)^{D}}\int_{-\infty}^{\infty}\frac{d\vec{k}}{\omega_{k}^{s-1}}\cos(\vec{k}\cdot (\vec{x}-\vec{x'}));
 \label{field2} \\
    \left[\hat{\varphi}(\vec{x},t),\hat{\varphi}^{\dagger}(\vec{x'},t) \right]=\frac{(\alpha-\beta)}{(2\pi)^{D}}\int_{-\infty}^{\infty}\frac{d\vec{k}}{\omega_{k}^{s}}\cos(\vec{k}\cdot (\vec{x}-\vec{x'}));
 \label{field21} \\
  \left[\hat{\pi}(\vec{x},t),\hat{\pi}^{\dagger}(\vec{x'},t) \right]=-\frac{(\alpha-\beta)}{(2\pi)^{D}}\int_{-\infty}^{\infty}\frac{d\vec{k}}{\omega_{k}^{s-2}}\cos(\vec{k}\cdot (\vec{x}-\vec{x'}));
 \label{field22} \\
    \left[\hat{\varphi}(\vec{x},t),\hat{\pi}^{\dagger}(\vec{x'},t) \right]=0;
\label{field23}
\end{eqnarray}
the last vanishing commutator is due to the trivial commutators $[\hat{a},\hat{a}]=0=[\hat{b},\hat{b}]$, and it corresponds to the commutator (\ref{field1});
furthermore, with the choice $s=1$, and the election $\alpha=1=\beta$, one defines the traditional scheme with $\left[\hat{\varphi},\hat{\pi}\right]=i \delta_{Dirac}$, and $\left[\hat{\varphi},\hat{\varphi}^{\dagger}\right]=0= \left[\hat{\pi},\hat{\pi}^{\dagger}\right]$; note that with this choice of parameters, one decides which commutator is nontrivial, and then the commutation relations (\ref{field})-(\ref{field1}) are reproduced. However, this is only an election, since one can for example to choice $\alpha=1=-\beta$, and thus $\left[\hat{\varphi},\hat{\pi}\right]=0$, and  $\left[\hat{\varphi},\hat{\varphi}^{\dagger}\right]\neq 0\neq \left[\hat{\pi},\hat{\pi}^{\dagger}\right]$, which corresponds certainly to an atypical version for quantum field theory. 

Therefore, if in general $\alpha\neq \pm \beta$, we have the more stringent version of the commutation relations
(\ref{field})-(\ref{field1}), which will be nonvanishing, close in spirit to noncommutative quantum mrchanics, in which the commutation relations (\ref{qm2}) are precisely nonvanishing \cite{falomir,aca,nair}. Hence, one can start with the initial idea of a transition to the continuum from the relations (\ref{qm1}), and (\ref{qm2}), to the relations (\ref{field})-(\ref{field0}), but one obtains at the end a more general version for quantum field theory, without invoking any breakthrough of the modern physics. In the same sense, if we look back, in the coverse direction of that transition, then one can infers that a more stringent version for quantum mechanics there exists, in which the coordinates and momenta do not commute to each other.

We discuss in detail only the fundamental commutator (\ref{field2}); independently on the choice of the parameters $\alpha$, and $\beta$, the integrals for the relations (\ref{field21}), and (\ref{field22}) can be developed along the same lines. Thus, in the expression (\ref{field2}),
the value $s=1$ corresponds to  the usual Dirac delta in $D$ dimensions; this general integral for $s$ arbitrary will define the distributions 
$ D_{s}$, that we shall determine explicitly in different dimensions. Note that the cases for $s<1$ are evidently divergent, and we work with the restriction $s\geq 1$, with the purpose of including the standard case $s=1$; as we shall see, the conventional description in terms of the Dirac delta distribution separates the infinite family  of divergent theories, from the infinite family of convergent ones constructed here. The k-integration
in the general expression (\ref{field2}) must be determined explicitly for each dimension, and in principle it exists in arbitrary spacetime dimensions.\\

\subsection{1+1 QFT}
\label{oneplusone}
Low dimensional quantum field theories appear naturally in condensed matter systems, and as toy models of realistic four dimensional theories.
We construct first the $1+1$ field theory, which will be compared with higher dimensional cases, since the present formulation is sensitive on the background dimension.

Using the integral representation of the modified Bessel function of the second kind of order $\nu$, $$ K_{\nu}(x)=\frac{2^{\nu}}{\sqrt{\pi}x^{\nu}}\Gamma\left(\nu+\frac{1}{2}\right)\int_{0}^{\infty}\frac{\cos(xt)}{(t^{2}+1)^{\nu+\frac{1}{2}}}dt,$$ 
 the solutions are written as
\begin{eqnarray}
    \left[\hat{\varphi}(x,t),\hat{\pi}(x',t) \right]= (\alpha+\beta)
    \begin{cases}
    i\delta(x-x') & s=1, \\
     \frac{i2^{1-\nu}}{\sqrt{\pi}\Gamma(\nu-\frac{1}{2})}\frac{\abs{\Delta x}^{\nu-1} }{m^{\nu-1}} K_{\nu-1}(m \abs{\Delta x}) & s=2\nu, \\
     -\left.\frac{1}{(\nu-1)!}\frac{d^{(\nu-1)}}{dk^{(\nu-1)}}\left(\frac{1}{(k+im)^{\nu}}e^{ik\abs{\Delta x}}\right)\right\vert_{k=im} & s=2\nu +1;
    \end{cases}
    \label{field3}
\end{eqnarray}
with $\nu=1,2,...$, and $\Delta x=x-x'$; these solutions are shown in the next table for some values of $s$, where the global factor $\alpha+\beta$ has been omitted by simplicity.
The maximum values occur at $\Delta x \rightarrow 0$, for a fixed $m$, are finite for $s\geq3$, and diverge logarithmically for $s=2$. In general the distributions  depend on the space-like separation ${x}-{x'}$, which is preserved under $1+1$ Lorentz transformations.

\begin{center}
\begin{tabular}{|c|c|c|}
\hline
     s &  $\left[\hat{\varphi},\hat{\pi}\right]$ & $\lim _{\Delta x \rightarrow 0} \left[\hat{\varphi},\hat{\pi} \right]$ \\ \hline
     1 &  $i\delta(\abs{\Delta x})$ & $i\infty$ \\ \hline
     2 & $\frac{i}{\pi} K_{0}(m \abs{\Delta x})$ & $-i\frac{1}{\pi}\ln\frac{m\abs{\Delta x}}{2}$  \\ \hline
     3 & $\frac{i}{2m}e^{-m\abs{\Delta x}}$ & $i\frac{1}{2m}$ \\ \hline
     4 & $i\frac{\abs{\Delta x}}{\pi m} K_{1}(m\abs{\Delta x})$ & $i\frac{1}{\pi m^{2}}$ \\ \hline
     5 & $i \frac{1}{4m^{3}}(1+m \abs{\Delta x})e^{-m\abs{\Delta x}}$ & $i\frac{1}{4m^{3}}$ \\ \hline
\end{tabular}
\end{center}

\begin{figure}[H]
  \begin{center}
   \includegraphics[width=.55\textwidth]{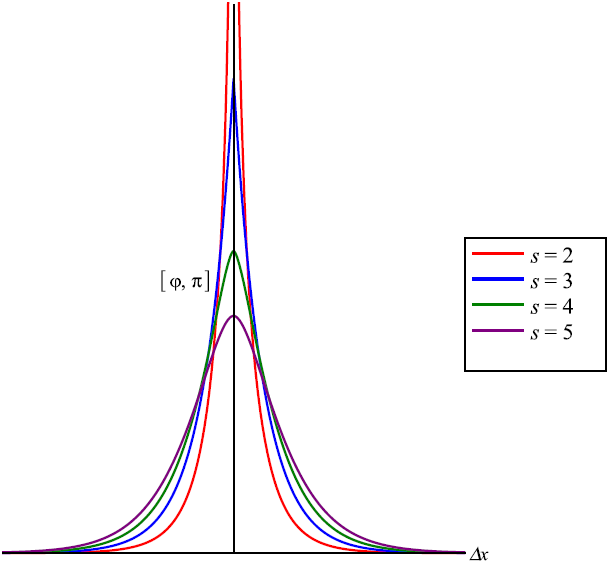}
  \caption{$\left[\hat{\varphi},\hat{\pi}\right]$ as function on $\Delta x$, for different values of $s$, and for $m>1$.}  
  \label{fig1}
\end{center}
\end{figure}
\begin{figure}[H]
  \begin{center}
   \includegraphics[width=.55\textwidth]{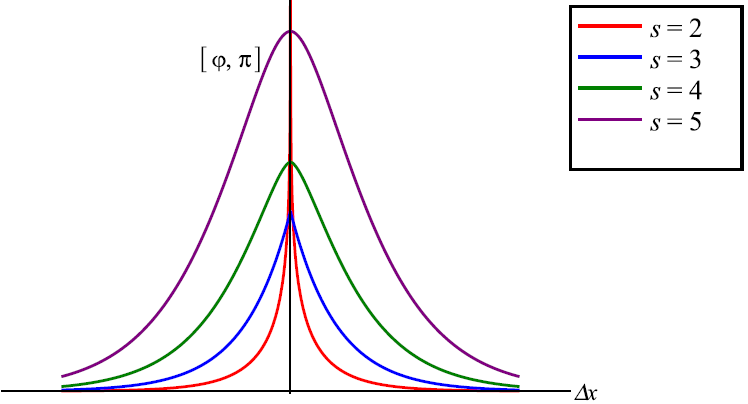}
  \caption{ $\left[\hat{\varphi},\hat{\pi}\right]$ as function on $\Delta x$ for different values of $s$, and with $m<1$.}  
  \label{fig2}
\end{center}
\end{figure}

The field commutator can be represented as function on $\Delta x$ for fixed $m$, and for the different values of $s$;
the maximums correspond to $\Delta x \rightarrow 0$, and according to the table, they are proportional to $\frac{\alpha+\beta}{m^{s-2}}$ for  $s\geq 3$; 
therefore, the maximums are decreasing as $s\rightarrow \infty$ for $m> 1$ (figure \ref{fig1}), and they are increasing as $s\rightarrow \infty$ for $m<1$ (figure \ref{fig2}); hence, the distribution of the curves illustrated in figure \ref{fig1} is inverted in the figure \ref{fig2}.

\subsection{2+1 QFT}
\label{twoplusone}
\label{twoone}
In the standard scheme it is assumed that the field commutator is determined by the Dirac delta in all dimensions; in the scheme at hand,
in two spatial dimensions we deal with integrals of (zero-order) Bessel functions of the first kind, by using polar coordinates and integrating the angular variables,
\begin{eqnarray}
    \left[\hat{\varphi}(\vec{x},t),\hat{\pi}(\vec{x'},t) \right]=(\alpha+\beta)\frac{i}{2\pi}\int_{0}^{\infty}dk\frac{k}{(\sqrt{k^{2}+m^{2}})^{s-1}}J_{0}(k\abs{\Delta \vec{x}}),
\label{2d}
\end{eqnarray}
where $k^2=k_{1}^2+k_{2}^{2}$; the next table shows explicit solutions of the integral (\ref{2d}) for some values of $s$  \cite{Tables}; this table must be compared directly with the previous table, particularly the appearance of the function $e^{-m\abs{\Delta \vec{x}}}$ and of the  modified Bessel function of the second kind. Note that in this case the commutator is finite at short distances for $s\geq 4$, and has basically the form $\frac{\alpha+\beta}{m^{s-3}}$; the logarithmic divergence appears for $s=3$.
\begin{center}
\begin{tabular}{|c|c|c|}
\hline
     s &  $\left[\hat{\varphi},\hat{\pi} \right]$ & $\lim _{\abs{\Delta \vec{x}} \rightarrow 0} \left[\hat{\varphi},\hat{\pi}\right]$ \\ \hline
     1 &  $i\frac{1}{2\pi}\frac{\delta(\abs{\Delta \vec{x}})}{\abs{\Delta \vec{x}}}$ & $i\infty$ \\ \hline
     2 & $\frac{i}{2\pi}\frac{1}{\abs{\Delta \vec{x}}}e^{-m\abs{\Delta \vec{x}}}$ & $\frac{i}{2\pi}\frac{1}{\abs{\Delta \vec{x}}}$  \\ \hline
     3 & $i\frac{1}{2\pi}K_{0}(m\abs{\Delta \vec{x}})$ & $-\frac{i}{2\pi}\ln\frac{m\abs{\Delta \vec{x}}}{2}$ \\ \hline
     4 & $i\frac{1}{2\pi m} e^{-m\abs{\Delta \vec{x}}}$ & $\frac{i}{2\pi}\frac{1}{m}$ \\ \hline
     5 & $i\frac{\abs{\Delta \vec{x}}}{4\pi m}K_{1}(m\abs{\Delta \vec{x}})$ & $\frac{i}{4\pi}\frac{1}{ m^{2}}$ \\ \hline
\end{tabular}
\end{center}
The expressions for higher values of $s$ can be computed by using the recurrence relation 
\begin{equation}\label{recurrence}
    f(s+2,m,\abs{\Delta \vec{x}})=\frac{1}{m(1-s)}\frac{\partial f(s,m,\abs{\Delta \vec{x}})}{\partial m}, \qquad s\geq 2,
\end{equation}
where $f(s,m,\abs{\Delta \vec{x}})$ is given by the integral in (\ref{2d}). For example, one can verify that the expression of the commutator for $s=4$ is obtained by using (\ref{recurrence}) with the expression for $s=2$; in the same way, in order to obtain the commutator for $s=6$, the relation (\ref{recurrence}) must be applied to $f(4,m,\abs{\Delta \vec{x}})$, and so on. \\
Despite the relation (\ref{recurrence}) holds for every $s\geq 2$, the expressions for odd values of $s$ can be computed directly thanks to the integral
\begin{equation}
\int_{0}^{\infty} \frac{J_{0}(b k)\  k}{\left(\sqrt{k^{2}+m^{2}}\right)^{s-1}} d x=\frac{b^{\frac{s-3}{2}}}{(2m)^{\frac{s-3}{2}} \Gamma(\frac{s-1}{2})} K_{\frac{s-3}{2}}(m b), \qquad \text{for $s$ odd.}
\end{equation}

\subsection{3+1 QFT}
\label{threeone}
On the other hand, in three spatial dimensions the commutator in spherical coordinates is written as
\begin{eqnarray}
    \left[\hat{\varphi}(\vec{x},t),\hat{\pi}(\vec{x'},t) \right]= (\alpha+\beta)\frac{i}{2\pi ^{2} \abs{\Delta \vec{x}}}\int_{0}^{\infty}dk \frac{k\sin{(k\abs{\Delta \vec{x}})}}{\left(\sqrt{k^{2}+m^{2}}\right)^{s-1}},
\label{3d}
\end{eqnarray}
where $k=\sqrt{k_{1}^2+k_{2}^{2}+k_{3}^{2}}$; some solutions are \cite{Tables},
\begin{center}
\begin{tabular}{|c|c|c|}
\hline
     s &  $\left[\hat{\varphi},\hat{\pi}\right]$ & $\lim _{\abs{\Delta \vec{x}} \rightarrow 0} \left[\hat{\varphi},\hat{\pi}\right]$ \\ \hline
     1 &  $i\frac{1}{2\pi}\frac{\delta(\abs{\Delta \vec{x}})}{\abs{\Delta \vec{x}}^{2}}$ & $i\infty$ \\ \hline
     2 & $i\frac{m}{2\pi^{2}\abs{\Delta \vec{x}}}K_{1}(m\abs{\Delta \vec{x}})$ & $\frac{i}{2\pi^{2}}\frac{1}{\abs{\Delta \vec{x}}^{2}}$ \\ \hline
     3 & $i\frac{1}{4\pi\abs{\Delta \vec{x}}}e^{-m\abs{\Delta \vec{x}}}$ & $\frac{i}{4\pi}\frac{1}{\abs{\Delta \vec{x}}}$  \\ \hline
     4 & $i\frac{1}{2\pi^{2}}K_{0}(m\abs{\Delta \vec{x}})$ & $-\frac{i}{2\pi^{2}}\ln\frac{m\abs{\Delta \vec{x}}}{2}$ \\ \hline
     5 & $i\frac{1}{8\pi m} e^{-m\abs{\Delta \vec{x}}}$ & $\frac{i}{8\pi}\frac{1}{m}$ \\ \hline
      6 & $\frac{i}{6\pi^{2}}\frac{\abs{\Delta \vec{x}}}{m}K_{1}(m\abs{\Delta \vec{x}})$ & $\frac{i}{6\pi^{2}}\frac{1}{m^{2}}$ \\ \hline
      \end{tabular}
\end{center}
the convergence at short distances takes the form $\lim _{\abs{\Delta \vec{x}} \rightarrow 0} \left[\hat{\varphi},\hat{\pi}\right]\sim\frac{\alpha+\beta}{m^{s-4}}$, with $s\geq 5$; in contrast with 
the previous cases such a convergence is achieved from $s\geq 4$, for two spatial dimension, and from $s\geq 3$ for one spatial dimension; in fact these results suggest that for a $D+1$ space-time dimension, the convergence is achieved for $s\geq D+2$; additionally the logarithmic divergence is achieved just when the weight coincides with the spacetime dimension, $s=D+1$. Such a coincidence allows, in particular, to choice the function $K_{0}$ as the common expression for the field commutator in all dimensions.
General expressions for the integrals (\ref{3d}) for every $s$ are available in the literature\cite{Tables}.\\

\section{Finite observables}
\label{vev}
We construct now the Hamiltonian, momentum and charge operators; we start from their classical form
\begin{subequations}
\begin{align}
    H &=\int d^{D} x\left(\pi_{\varphi}^{*} \pi_{\varphi}+\nabla \varphi^{*} \cdot \nabla \varphi+m^{2} \varphi^{*} \varphi\right),
\label{hami}\\
    Q &=i\int d^{D}x\ (\varphi^{\dagger}\partial_{0}\varphi-\partial_{0}\varphi^{\dagger}\varphi), \\
    P_{i} &=-i\int d^{D}x\ (\pi_{\varphi}\partial_{i}\varphi+\pi_{\varphi^{\dagger}}\partial_{i}\varphi^{\dagger});
\end{align}
\label{class}
\end{subequations}
as it is well known, the quantization ambiguities imply that one has that to choice the order in which the commuting classical quantities will be promoted to operators; for example, if one considers  {\it symmetrized}
classical expressions such as $\pi_{\varphi}^{*} \pi_{\varphi} = \frac{1}{2} \pi_{\varphi}^{*} \pi_{\varphi} +\frac{1}{2} \pi_{\varphi}\pi_{\varphi}^{*}$, then the operator versions of the observables read
\begin{subequations}\label{operators}
\begin{align}
    \hat{H}&= 2\int d^{D}k\ \omega_{k}^{2-s}\left[\hat{a}^{\dagger}(\vec{k})\hat{a}(\vec{k})+\hat{b}^{\dagger}(\vec{k})\hat{b}(\vec{k}) \right]+\underbrace{ (\alpha+\beta)\frac{L^{D}}{(2\pi)^{D}}\int d^{D}k\ \omega_{k}^{2-s}}\\
    \hat{Q} &= 2\int d^{D}k\ \omega_{k}^{1-s}\left[\hat{a}^{\dagger}(\vec{k})\hat{a}(\vec{k})-\hat{b}^{\dagger}(\vec{k})\hat{b}(\vec{k}) \right] +\underbrace{ (\alpha-\beta)\frac{L^{D}}{(2\pi)^{D}}\int d^{D}k\ \omega_{k}^{1-s}}, \\
    \hat{P}_{i}&=2\int d^{D}k\ \frac{k_{i}}{\omega_{k}^{s-1}}\left[\hat{a}^{\dagger}(\vec{k})\hat{a}(\vec{k})+\hat{b}^{\dagger}(\vec{k})\hat{b}(\vec{k}) \right]+ (\alpha+\beta)\frac{L^{D}}{(2\pi)^{D}}\underbrace{\int d^{D}k\ k_{i}\ \omega_{k}^{1-s}}_{=0}.
\label{momenta}
\end{align}
\end{subequations}
where we have confined the system in a box with sides of length $L$. The above expressions have been obtained by using the commutators (\ref{ac}), by locating the annihilation operators to the right hand side, without invoking normal ordering; in the approach at hand, the potentially divergent integrals can be controlled by chossing appropriately the weight $s$, and then they will be finite.

With the usual definition of the vacuum, $\hat{a}(k)\ket{0} =0= b(k) \ket{0}$, the first terms with the annihilation operators located to the right hand side vanish trivially. Therefore,
the action of the observables on the vacuum state is determined by the terms in underbrace, which are potentially divergent, depending on the choice of the parameters $\alpha$, $\beta$, and the weight $s$; for  the momenta (\ref{momenta})
such an integral vanishes trivially for all dimensions, because the integrand is an odd function on $k_{i}$. Furthermore, the choice  $\alpha= \beta$ will lead a neutral vacuum, and to a nontrivial vacuum energy; conversely, the choice $\alpha= -\beta$ will lead to a vanishing vacuum energy, and to a nontrivial vacuum charge. Hence, if one imposses the constraint  $\alpha\neq \pm \beta$, then the vacuum will have both a nontrivial energy and a nontrivial charge; however in the present approach both quantities can be regularized, removing the UV divergences.

On the other side, if the transition to operators is made without using the {\it symmetrization} of classical expressions, but we use for example the expressions such as Eqs. (\ref{class}), then the coefficients in the integrals in underbrace are simply $\alpha$, or $\beta$; however, the key observation in the approach at hand is that such integrals are finite, and the normal ordering of operators is not invoked.

\subsection{The vacuum energy is finite}
\label{ve}
This case includes the traditional values $\alpha=1=\beta$, that together with the 
 choice $s=1$ lead to  a neutral vacuum state $\hat{Q}\ket{0}=0$, and with ultraviolet  divergences for $\hat{H}$; 
 this represents the first famous result of QFT, an infinite energy for the vacuum, which is removed by invoking normal ordering of operators; as we shall see, this procedure is not required in the present scheme.

In the approach at hand the zero-point energy reads,
\begin{equation}
    \hat{H}\ket{0}=\frac{(\alpha+\beta) L^{D}}{(2\pi)^{D}} \int \ \frac{d^{D}k}{\left(\sqrt{|\vec{k}|^{2}+m^{2}}\right)^{s-2}}\ket{0}\equiv\frac{(\alpha+\beta) L^{D}}{(2\pi)^{D}} H_{0}\ket{0}.
\end{equation}
in the next table we show some values for the integral $H_{0}$; 
\begin{center}
\begin{tabular}{|c|c|c|c|}
\hline
     s &  $H_{0,D=1}$ & $H_{0,D=2}$ & $H_{0,D=3}$ \\ \hline
     1 &  $\infty$ & $\infty$ & $\infty$     \\ \hline
     2 &  $\infty$ & $\infty$ & $\infty$ \\ \hline
     3 &  $ln\frac{k}{m}$ & $\infty$ & $\infty$ \\ \hline
     4 & $\frac{\pi}{m}$ &  $ln\frac{|\vec{k}|}{m}$ & $\infty$\\ \hline
     5 & $\frac{2}{m^{2}}$ & $\frac{2\pi}{m}$ &  $ln\frac{|\vec{k}|}{m}$  \\ \hline
     6 & $\frac{\pi}{2}\frac{1}{m^{3}}$ & $\frac{\pi}{m^{2}}$ & $\frac{\pi^2}{m}$ \\ \hline
\end{tabular}
\end{center}
the first row  with $s=1$ is justly the traditional divergent case, valid for all dimensions; this result is replicated for the case $s=2$. In general for  $D+1$ space-time dimensions, the vacuum energy is finite from $s\geq D+3$, and takes the form $\frac{\alpha+\beta}{m^{s-D-2}}$; therefore, one can extend the table to the right and below, since the diagonals have basically the same form. In particular the diagonal with 
logarithmic divergences is obtained with the approximation $\frac{m}{k}<<1$ as $k\rightarrow \infty$ after the integration. Therefore, the general conclusion is that one choices whether the vacuum has or not an infinite energy; for a spacetime dimension given, there exist in fact an infinite number of values for $s$ leading to finite values for the vacuum energy; see the Eqs. (\ref{strong}), and (\ref{ew}) below for some numerical examples.

Clasically the potential for the complex field $\varphi$ coming from the Lagrangian defines a paraboloid, namely $V(\varphi,\bar{\varphi})=m^2\varphi\bar{\varphi}=m^2(\varphi^2_{1}+\varphi^2_{2})$, whose lowest energy level implies that $\varphi=0$, {\it i.e.} the bottom
of such a paraboloid. According to the table, the vev for the energy does not vanish (and in general does not diverge), rather it is finite and is defined in terms of the mass. This result can be considered as the analogous of the well known result for a quantum harmonic oscillator, whose potential energy is defined as $V(x)= \frac{1}{2}\omega^2 x^2$; clasically the vacuum is the state in which the particle is motionless, with $x=0$; however, quantum-mechanically the lowest energy state has an energy $E_{0}=\frac{1}{2} \hbar \omega$. This zero-point energy was fundamental for recovering, at the Einstein time, the expected classical limit for the average energy of an oscillator in thermal equilibrium at temperature $T$; especifically such a limit is obtained by expanding the Plack formulae $E_{w}=\frac{\hbar\omega}{e^{\hbar\omega/kT}-1}+\frac{1}{2}\hbar\omega$, in the classical limit $kT\gg \hbar \omega$; for more details on this issue and its relationship with the cosmological constant problem see \cite{sola}. As well known, there no exists in the traditional scheme, an anologous result due to the divergence of the vev for the energy;
furthermore, as we shall see in the section \ref{fluctuation},  
the vev for the quantum field itself, will not diverge.

Futhermore, for the excited states of the quantum harmonic oscillator, obtained from the repeated action of the creation operator on the vacuum,  $\ket{n}\equiv (\hat{a})^{\dagger}\ket{0}$, we have the very known eigen-energy expression $\hat{H}\ket{n}=(n+\frac{1}{2})\hbar \omega \ket{n}$, and then the system has a ladder of energy states. In our case, we have a similar situation, by considering the excitations of the field; if $\hat{a}^{\dagger}\ket{0}\equiv\ket{k_{a}}$ and $\hat{b}^{\dagger}\ket{0}=\ket{k_{b}}$ are single excited states, then it is straightforward to construct the following energy eigenstates,
\begin{eqnarray}
\hat{H}\ket{k_{a}}=\big[ \alpha\frac{2}{\omega_{k_{a}}^{s-2}}+(\alpha+\beta)\frac{L^{D}}{(2\pi)^D}H_{0}\big]\ket{k_{a}}, \nonumber\\
\hat{H}\ket{k_{b}}=\big[\beta \frac{2}{\omega_{k_{b}}^{s-2}}+(\alpha+\beta)\frac{L^{D}}{(2\pi)^D}H_{0}\big]\ket{k_{b}};
\end{eqnarray}
where the vacuum energy $H_{0}=H_{0}(m;s)$ is described above in the table; thus, these one-particle states have excitation energies $\frac{2\alpha}{\omega_{k_{a}}^{s-2}}$, and $\frac{2\beta}{\omega_{k_{b}}^{s-2}}$ in relation to the vacuum energy. These single expressions can be generalized for multi-particle eigen-states with $n$ a-bosons excited, and with $m$ b-bosons excited, in any order,
\begin{eqnarray}
\hat{H}\ket{ k_{a}^{1},,,k_{a}^n; k_{b}^{1},,,k_{b}^m}=\Big[ 2\alpha \sum_{i=1}^{n}(\frac{1}{\omega_{k_{a}^{i}}})^{s-2}+ 2\beta \sum_{i=1}^{m}(\frac{1}{\omega_{k_{b}^{i}}})^{s-2} +(\alpha+\beta)\frac{L^{D}}{(2\pi)^D}H_{0}\Big] \ket{ k_{a}^{1},,,k_{a}^n; k_{b}^{1},,,k_{b}^m};
\label{multi}
\end{eqnarray}
hence, the energy of this eigen-state is given by the sum of the energies of the various particles.
Additionally these energy eigen-states are also eigen-states for the number operators, which are defined as usual,
\begin{eqnarray}
\int d\vec{k} \hat{a}^{\dagger}(\vec{k})\hat{a}(\vec{k}) \ket{ k_{a}^{1},,,k_{a}^n; k_{b}^{1},,,k_{b}^m}=n\alpha \ket{ k_{a}^{1},,,k_{a}^n; k_{b}^{1},,,k_{b}^m}, \nonumber\\
\int d\vec{k} \hat{b}^{\dagger}(\vec{k})\hat{b}(\vec{k}) \ket{ k_{a}^{1},,,k_{a}^n; k_{b}^{1},,,k_{b}^m}=m\beta \ket{ k_{a}^{1},,,k_{a}^n; k_{b}^{1},,,k_{b}^m};
\end{eqnarray}
which count effectively the number of particles of each type in the multi-particle states.

In the traditional scheme the normal ordering implies  to forget (for a moment) the fundamental commutators (\ref{ac}) for extracting an infinite vacuum energy; however, in order to construct a nontrivial quantum field theory, one must to restore those commutators, and then one does not use normal ordering for constructing the eigenstates for the Hamiltonian and for the number operators. This ambiguity has darkened the traditional formulation of quantum field theory, as opposed to the approach at hand, in which the fundamental commutators (\ref{ac}) are maintained always switched on, namely, for constructing finite vev for the obsevables, and for the building of the multi-particle eigen-states described above.

\section{The vev for the field operator is finite}
\label{fluctuation}

One consequence of the usual choice $s=1$ is the inexistence of normalizable states; with the conventional definition of the vacuum state $\hat{a}|0>=0=\hat{b}|0>$, one has that, 
$<0|\hat{\psi}\hat{\psi}^{\dagger}|0>\rightarrow \infty$;
hence, the action of the field operators on the Hilbert space is not well defined. Since operators and expectation values normalize to delta functions in the usual formulation, one requires to construct well defined operators by smearing 
those singular distributions, which can be achieved by creating a wave-packet through
$ \int d \vec{p}e^{-ip\cdot x}f(p)\hat{a}^{\dagger}
|0>$, where typically the smearing function $f$ is chosen as the Gaussian $f=e^{-p^2/2m^2}$; this procedure is of course as arbitrary as the smearing functions chosen, and this kind of arbitrariness also has darkened the traditional formulation of quantum field theory.

However, as expected at this point, the fluctuation of the field operator at a fixed point is in general finite for arbitrary weight $s$; explicitly we have for arbitrary dimension,
\begin{eqnarray}
<0|\hat{\varphi}\hat{\varphi}^{\dagger}|0>=\frac{<0|0>}{(2\pi)^D}\alpha\int^{+\infty}_{-\infty} \frac{d\vec{k}}{w_{k}^{s}};
\quad
<0|\hat{\varphi}^{\dagger}\hat{\varphi}|0>=\frac{<0|0>}{(2\pi)^D}\beta\int^{+\infty}_{-\infty} \frac{d\vec{k}}{w_{k}^{s}};
\label{smear}
\end{eqnarray}
therefore,
\begin{equation}
\begin{aligned}[b]
\bra{0}\left(\hat{\varphi}(\textbf{x})\hat{\varphi}^{\dagger}(\textbf{x})+\hat{\varphi}^{\dagger}(\textbf{x})\hat{\varphi}(\textbf{x})\right)\ket{0} &=\frac{<0|0>}{(2\pi)^D}(\alpha+\beta)\int^{+\infty}_{-\infty} \frac{d\vec{k}}{w_{k}^{s}}\\ 
\end{aligned}
\label{smear1}
\end{equation}
the integral is shown in the next table for some values of $s$.
\begin{center}
\begin{tabular}{|c|c|c|c|}
    \hline
    $\left.\right.$ & \multicolumn{3}{|c|}{ $\int d\vec{k}{w_{k}^{-s}} $} \\ \hline
    $s$ & $D=1$ & $D=2$ & $D=3$  \\ \hline
    1   & $\infty$ as $ln(\frac{k}{m})$  & $\infty$ & $\infty$ \\ \hline
    2   & $\frac{\pi}{m}$ & $\infty$ as $ln(\frac{k}{m})$ & $\infty$ \\ \hline
    3   & $\frac{2}{m^{2}}$ & $\frac{2\pi}{m}$ & $\infty$ as $ln(\frac{k}{m})$ \\ \hline
    4   & $\frac{\pi}{2}\frac{1}{m^{3}}$ & $\frac{\pi}{m^{2}}$ & $\frac{\pi^{2}}{m}$ \\ \hline
    5 & $\frac{4}{3}\frac{1}{m^{4}}$ & $\frac{2}{3}\frac{\pi}{m^{3}}$ & $\frac{4}{3}\frac{\pi}{m^{2}}$ \\ \hline
    6 & $\frac{3}{8}\frac{\pi}{m^{5}}$ & $\frac{\pi}{2}\frac{1}{m^{4}}$ & $\frac{\pi^{2}}{4}\frac{1}{m^{3}}$ \\ \hline
\end{tabular}
\end{center}

\section{The case $\alpha=-\beta$}
This simple algebraic constraint eliminates automatically the infrared and ultraviolet divergences for the vacuum energy, although now the vacuum charge is susceptible of suffering such divergences; for this case one can construct a table similar  to the table described above for the vacuum enegy, by considering the integral in the expression (\ref{operators}) for the charge operator. For this atypical case the field commutators will  read
$\left[\hat{\varphi},\hat{\pi}\right]=0$, and  $\left[\hat{\varphi},\hat{\varphi}^{\dagger}\right]\neq 0\neq \left[\hat{\pi},\hat{\pi}^{\dagger}\right]$, according to the expressions (\ref{field2})--(\ref{field22}); additionally the vacuum expectation values fot the field operator vanish trivially under this election of parameters (see Eq.(\ref{smear1}) above).
This QFT has not the wanted structure, since one is expecting a nontrivial field commutator $\left[\hat{\varphi},\hat{\pi}\right]$; however, it illustrates the sensibility of the resulting QFT on the choice of the parameters $\alpha$, and $\beta$.

\section{On noncommutative field theory}

The noncommutative version for the quantum mechanics of a harmonic oscillator, can be defined by the following commutation relations for the dynamical variables,
\begin{eqnarray}
[{q_{i},p_{j}}]=i \delta_{ij};\quad
[{q_{1},q_{2}}]=i \theta, \quad [{p_{1},p_{2}}]=i B;
\end{eqnarray}
inspired by this noncommutative scheme, in \cite{gamboa} a noncommutative field theory is introduced by considering the transition to the continuum through the following 
commutation relations for a complex scalar field,
\begin{eqnarray}
[\hat{\psi}(\vec{x}), \hat{\Pi}_{\psi}(\vec{y})]=i\delta(\vec{x}-\vec{y}); \quad
\label{ncf}\\
\big[\hat{\psi}(\vec{x}), \hat{\psi}^{\dagger}(\vec{y})\big]= \theta\delta(\vec{x}-\vec{y}), \quad
\big[\hat{\Pi}_{\psi}(\vec{x}), \hat{\Pi}_{\psi}^{\dagger}(\vec{y})\big]= B\delta(\vec{x}-\vec{y});
\label{ncf1}\\
\big[\hat{\psi}(\vec{x}), \hat{\Pi}_{\psi}^{\dagger}(\vec{y})\big]=0; \label{ncf2} 
\end{eqnarray}
which correspond to a generalization of the commutation relations (\ref{field})-(\ref{field1}).  There is no a comment on the commutator (\ref{ncf2}) in \cite{gamboa}, and we assume here that it vanishes in that approach; this vanishing commutator coincides with our commutator (\ref{field23}), which is a consequence of the trivial commutators $[\hat{a},\hat{a}]=0=[\hat{b},\hat{b}]$. In the approach \cite{gamboa}, the parameters $\theta$, and $B$ measure the noncommutativity, and their phenomenological bounds are established; such bounds do not contradict the low-energy phenomena, and will have observable effects at high energies.

According to the Eqs. (\ref{field2})--(\ref{field23}) of the present approach, the more stringent version of the noncommutative quantum field theory can be obtained with the restriction $\alpha\neq \pm\beta$. Since each commutator is defined with a different weight $s$, one can not to identify all commutators with the Dirac delta, such as in the above proposal; one can identify only one commutator with that distribution, for example with $s=2$ the commutator (\ref{field22}) is fixed as the Dirac delta, and the rest of the commutators can be determined as already explained previously,
\begin{eqnarray}
    \left[\hat{\varphi}(\vec{x},t),\hat{\pi}(\vec{x'},t) \right]=\frac{i(\alpha+\beta)}{2\pi^{2}}\frac{m}{|\Delta \vec{x}|}K_{1}(m|\Delta\vec{x}|);
 \label{ap2} \\
    \left[\hat{\varphi}(\vec{x},t),\hat{\varphi}^{\dagger}(\vec{x'},t) \right]=(\alpha-\beta)\frac{e^{-m|\Delta\vec{x}|}}{4\pi |\Delta\vec{x}|};
 \label{ap21} \\
  \left[\hat{\pi}(\vec{x},t),\hat{\pi}^{\dagger}(\vec{x'},t) \right]=-(\alpha-\beta)\delta  (\vec{x}-\vec{x'});
 \label{ap22} \\
    \left[\hat{\varphi}(\vec{x},t),\hat{\pi}^{\dagger}(\vec{x'},t) \right]=0.
\label{ap23}
\end{eqnarray}
However, this choice for $s$ is only for comparative reasons, since the obsevables such the vacuum energy diverge for $s=2$ in four dimensional space-time. Therefore, according to the previous results, in a four dimensional space-time, the first value for the weight $s=6$ guarantees the convergence for the vacuum energy and for the vev"s of the field operators, which are of the form $\frac{\alpha+\beta}{m}$, and $\frac{\alpha+\beta}{m^3}$ respectively; moreover, for the field commutators we have,
\begin{eqnarray}
    \left[\hat{\varphi}(\vec{x},t),\hat{\pi}(\vec{x'},t) \right]=i(\alpha+\beta)\frac{|\Delta \vec{x}|}{6 \pi^2 m}K_{1}(m|\Delta\vec{x}|);
 \label{aap2} \\
    \left[\hat{\varphi}(\vec{x},t),\hat{\varphi}^{\dagger}(\vec{x'},t) \right]=(\alpha-\beta)\frac{m|\Delta\vec{x}|+1}{32\pi m^3} e^{-m|\Delta\vec{x}|};
 \label{aap21} \\
  \left[\hat{\pi}(\vec{x},t),\hat{\pi}^{\dagger}(\vec{x'},t) \right]=-(\alpha-\beta) \frac{e^{-m|\Delta\vec{x}|}}{8\pi m};
 \label{aap22} \\
    \left[\hat{\varphi}(\vec{x},t),\hat{\pi}^{\dagger}(\vec{x'},t) \right]=0.
\label{aap23}
\end{eqnarray}
in fact, from $s\geq 6$ there exists an infinite family of (finite) noncommutative quantum field theories in the more stringent version that can represent an altervative for the proposal given in \cite{gamboa}; we shall study in forthcoming works the phenomenological impications of our approach along the lines developed in that reference.

\section{Particle physics numerology}
In quantum field theory
the energy scale is introduced through the commutators (\ref{ac}), and in the standard scheme, with the choice $\alpha=\hbar=\beta$, the fundamental commutator reads $   \left[\hat{\varphi},\hat{\pi} \right] \sim\hbar \delta_{Dirac}$. In the approach at hand,  the parameters $\alpha$ and $\beta$ can be chosen independently, and in principle they define different energy scales; for simplicity we consider here that they define the same energy scale.
In the system of units in which all fundamental quantities appear explicitly, we have that $[\alpha]=[\beta]= [\hbar]/ [T]^{s-1}$, and hence these parameters will be adjusted according to the value of $s$; the choice $s=1$ reproduces the traditional case. Therefore, we have a new timelike parameter encodes in $\alpha$ and $\beta$; this parameter can be identified with the corresponding time scale associated with the energy scale, and particularly it can be identified with the Planck time. If
we call $T$ such a parameter, then the vacuum energy can be rewritten
as (neglecting factors like $2\pi,\pi^2$, etc., that appear in the tables),
\begin{eqnarray}
H_{0}(T,s)=\frac{mc^2}{(\frac{T}{\hbar}mc^2)^{s-2}};
\label{ppn1}
\end{eqnarray}
where we have considered that in the new system of units $m\rightarrow \frac{mc}{\hbar}$ and $w_{k}\rightarrow cw_{k}$; note that the quantity $\frac{T}{\hbar}mc^2 $ is dimensionless, and thus this expression is easily interpretable, since the vacuum energy turns out to be proportional to the rest energy of the field. Moreover, note that at quantum level, the particle has an {\it effective} mass $m_{_{effec}}\equiv  (\frac{T}{\hbar}mc^2)^{2-s} m$, which is different to that initially postulated at Lagrangian level.

Let $\frac{T}{\hbar}c^2 \equiv\frac{1}{M_{}}$ be the inverse of the mass scale defined by $T$, then 
\begin{eqnarray}
H_{0}(
M,s)=\frac{mc^2}{(\frac{m}{M})^{s-2}};
\label{ppn1}
\end{eqnarray}
therefore for masses above the mass scale $m>M$, the vacuum energy satisfies $H_{0}<mc^2$, and conversely for the case $m<M$, one has that $H_{0}>mc^2$; thus, in relation to the mass scale, the light masses will have a higher vacuum energy than those with heavy masses. In particular if $m=M$, then the vacuum energy will reduce to the energy that defines the scale, $H_{0}=Mc^2$.
This result is valid without reference to a mass scale, since we can compare the vacuum energy for two masses $m_{1}$, and $m_{2}$ (and for the same weight $s$),
\begin{eqnarray}
\frac{H_{0}(m_{2},s)} {H_{0}(m_{1},s)}=\Big(\frac{m_{1}}{m_{2}}\Big)^{s-3};
\end{eqnarray}
thus, if $m_{1}<m_{2}$, then $H_0(m_{1})>H_{0}(m_{2})$.

The present scheme can be applied to estimate the vev for the energy of a neutral Higgs boson-like particle, which is realized by imposing that  the field operator is Hermitian, thus $\hat{b}=\hat{a}$, and a vanishing charge is achieved by choosing $\alpha=\beta$.

First we choice as scenario the energy scale for the strong interactions with $M\approx 1 Gev$, and for the mass of the Higgs-like field the value determined experimentally for the Higgs boson, $m\approx 125 Gev$; therefore, from Eq. (\ref{ppn1}),
\begin{equation}
    \frac{H_{0}}{c^{2}}\approx (125)^{3-s}\ GeV \approx \begin{cases}
    15\ TeV, & \quad s=1,\\
    125\ GeV, & \quad s=2, \\
    1\ GeV, & \quad s=3, \\
    8\ MeV, & \quad s=4, \\
    60\ KeV, & \quad s=5, \\
    500\ eV, & \quad s=6, 
\label{strong}
    \end{cases}
\end{equation}
since that $m>M$, the energy is decreasing as $s\rightarrow\infty$; thus, if the value of this observable is determined experimentally, and such a value is below the mass of the Higgs-like particle, then one can to choice the appropriate value for $s$, and thus to set the corresponding quantum field theory. Note that, since there exist orders of magnitude between the values of the vev's for the different values of $s$, the adjustment of $s$
may require a fine tuning by using fractional values.

Another scenario of interest is 
the electroweak energy scale with $M\approx 246 Gev$,
 and hence we have the following values for the vev's of the energy for same Higgs-like particle of mass $m\approx 125 Gev$, in order to compare with the previous table of values; in this case the vev's are above the mass $m$, and they are increasing as $s\rightarrow \infty$;
\begin{equation}
    \frac{H_{0}}{c^{2}}\approx 125(2)^{s-2}\ GeV \approx \begin{cases}
    62\ GeV, & \quad s=1, \\
    125\ GeV, & \quad s=2, \\
    250\ GeV, & \quad s=3, \\
    500\ GeV, & \quad s=4, \\
    1\ TeV, & \quad s=5, \\
    2\ TeV, & \quad s=6. \\
    \end{cases}
\label{ew}
\end{equation}
Note that between the first three values for the vev's in the table (\ref{strong}), there exist orders of magnitude of difference, as opposed to the four first values in the table (\ref{ew}), which are approximately of the same order; along the same lines, one can compare the set of the vev's of the
last three values in both tables. Similarly one can compare the vev's in both tables for  the value $s=6$, which applies for the four-dimensional case as the first value for finite observables; we realize that there exist $10^{10}$ orders of magnitude of difference. Such a sensibility of the quantum field theory respect to the choice of the parameter $s$ will be relevant in the discussion below on the cosmological constant problem.

\section{On normal ordering and general relativity}
In the traditional scheme the basic physical argument for discarding the zero-point energy is that one is measuring only the energy differences; however, as well known, this idea is not valid in general relativity, since any form of energy is a source of gravity, and it is the whole of the energy of a system what contributes to the gravitational interaction, and hence, one must to renounce to normal ordering on a curved background. Furthermore, although the vacuum energy can be unobservable in a flat background, a charged vacuum can have observable effects; therefore one can not use normal ordering on the same grounds. Since the approach at hand allows us to control the UV divergences by using the weighted invariant measures, without invoking normal ordering, is consistent with general relativity; in the same sense, 
it is relevant that the vacuum energy is determined precisely by the (effective) mass of the field; note also that the vacuum energy can not to be fixed to zero within the present approach, unless one choices the constraint $\alpha=-\beta$.

\section{Quantum field theory and the cosmological constant problem}
As well known the cosmological constant problem arises in the intersection of quantum field theory and general relativity since the zero-point fluctuations of the quantum fields provide a nontrivial contribution to the vacuum energy density in the universe, which acts as a source to the cosmological constant that appears in the Einstein equations. The problem
arises due to the large discrepancy between the enormous vacuum energy density of various contributions from quantum field theory and the tiny bounds to the cosmological constant coming from large scale cosmology
(see \cite{hp}, and references cited therein, for a historical review).

Historical antecedents that shown a nontrivial link between quantum field theory and general relativity are in order;
W. Pauli made a computation that shown that if the zero-point energies of QED account for the energy density of the vacuum in gravitational backgrounds, then the radius of the universe is about 31 km;
this old scheme required, as an UV cut-off, the electron radius \cite{hp},\cite{stra}. Furthermore,  in the first published works on the contributions of the quantum fluctuations to the cosmological constant, 
Zel'dovich emphasized that the estimates of the vacuum energies of particle physics exceed the observational bounds for the cosmological constant in 46 orders of magnitude, which pointed to a fundamental problem in modern physics\cite{zel}. 

We have at hand a new approach to address the problem of the cosmological constant from a new perspective; first, the present scheme does not require an UV cut off, since it provides finite results for the vev's of the different observables. Second, we have at hand a free adjustable parameter $s$, which can lead to very large or very small vev's for the observables; since this parameter is susceptible to a fine tuning, then one may achieve concordance  between  the zero-point energy density with the observacional bounds of the cosmological constant; this will be subject of forthcoming works.\\

{\bf Acknowledgements:}
This work was supported by the Sistema Nacional de Investigadores (M\'exico). We also thank to the Consejo Nacional de Ciencia y Tecnología (CONACYT) for giving financial support to Amadeo Méndez Ugalde (CVU: 1007917, 2019-2021).

\end{document}